# Enhancement in Li-ion Conductivity through Co- doping of Ge and Ta in Garnet $Li_7La_3Zr_2O_{12}$ Solid Electrolyte

**Muktai Aote[a], A.V.Deshpande[a,*]**

[a] Department of Physics, Visvesvaraya National Institute of Technology, South Ambazari Road Nagpur, Maharashtra 440010, India

[*] **Corresponding Author:**

**Dr.(Mrs.)A. V. Deshpande**

Department of Physics,

Visvesvaraya National Institute of Technology,

South Ambazari Road, Nagpur

Maharashtra, 440010 (India)

E-mail Address: avdeshpande@phy.vnit.ac.in

Ph.No.: +91-712-280-1251

## Abstract

For being used as an electrolyte in All Solid State Batteries (ASSB), a solid electrolyte must possess ionic conductivity comparable to that of conventional liquid electrolytes. To achieve this conductivity range, the series $Li_{6.8-y}Ge_{0.05}La_3Zr_{2-y}Ta_yO_{12}$ (y = 0, 0.15, 0.25, 0.35, 0.45) has been synthesized using solid-state reaction method and studied using various characterization techniques. The highly conducting cubic phase is confirmed from XRD analysis. Structural information was collected using SEM and density measurements. The prepared ceramic sample containing 0.25 Ta, sintered at $1050^0$ C for 7.30 hrs shows the maximum ionic conductivityof 6.61 x $10^{-4}$ S/cm at $25^0$C. The air stability of the same ceramic has also been evaluated after exposure for 5 months. The minimum activation energy associated with the maximum conductivity of 0.25 Ta is 0.25 eV. The DC conductivity measurements were done to confirm the ionic nature of conductivity for all ceramic samples.The stable result of ionic conductivity makes the 0.25 Ta containing ceramic sample a promising candidate for solid electrolytes for ASSB applications.

## 1. Introduction

In the past few years, there has been a rapid increase in the demand for lithium batteries after the significant development in the electric vehicle industry. Its high energy density and reliable portability play an essential role in driving the researcher's attention to it [1]. Currently, liquid electrolytes are used in these batteries. These are constructed using lithium salt, organic solvent, and a diaphragm. Although these electrolytes help to achieve higher ionic conductivity, the major concern related to such liquid electrolytes is the formation of dendrite after the number of duty cycles, which eventually leads to short-circuiting, flammability, and leakage of the electrolytes [2–4]. This leads to a major safety concern for all electric vehicle customers [5]. Thus, one common and most important need emerged: the search for an alternative option to conventional liquid electrolytes. Here, a solid electrolyte being an answer to these problems plays an important role. The solid electrolyte is the essential key factor in all-solid-state batteries (ASSB) with higher safety grades which can promisingly replace the liquid electrolytes. But, on a large-scale application basis like liquid electrolytes, solid electrolytes still face many challenges associated with the lithium (Li) ion conductivity, chemical, and electrochemical stabilities[3,6], and many other factors which are barriers to its way of practical use[7–12].

Hence, the focus has been shifted towards finding possible ways to minimize these issues related to solid electrolytes and bringing them to the frontier in this field.

Various solid electrolytes have been studied to date, such as inorganic ceramic electrolytes[13–16], organic polymer electrolytes [17], and sulfide solid electrolytes [18]. Polymer solid electrolytes have been studied widely due to their easy synthesis process. But the ionic conductivity of these electrolytes is in the order of $10^{-5}$ S/cm – $10^{-6}$ S/cm at ambient temperature, which is very low compared to conventional liquid electrolytes ($10^{-2}$ S/cm – $10^{-3}$ S/cm). On the other hand, sulfide electrolytes possess high ionic conductivity, but their electrochemical stability with the Li anode is significantly less [19]. Whereas, $Li_3PO_4$ has high electrochemical strength but low ionic conductivity [20]. Also, in the development of ASSBs, electrolytes in the form of thin film have been fabricated for thin film batteries. LIPON is one of the examples of such thin film solid electrolyte. But, its lower conductivity value obstructs its path of being used in practical applications [21–23]. Likewise, many other solid electrolytes such as, $Li_{0.35}La_{0.55}TiO_3$ (LTT) structured as crystalline perovskite, NASICON type $Li_{1.4}Al_{0.4}Ti_{1.6}(PO_4)_3$ (LATP), oxide electrolytes such as (Li, La)$ZrO_2$, (Li, La)$TiO_3$, garnet structured $Li_7La_3Zr_2O_{12}$ (LLZO), $Li_{1.5}Al_{0.5}Ge_{1.5}(PO_4)_3$ (LAGP) and $Li_2S$-$P_2S_5$ which is superionic sulfide have been examined. Among all these categories of solid electrolytes, the oxide type has benefits over other solid electrolytes [24]. Thus more detailed investigation of its properties, structure, and conduction mechanism is needed.

To be used as an electrolyte in ASSB's, oxide electrolytes must have some properties mentioned as follows [24]:

1) Room temperature Li ion conductivity equal to or greater than $10^{-4}$ S/cm.

2) High electrochemical potential window with redox stability of lithium.

3) Lower sintering temperature to minimize lithium loss and interfacial resistance.

4) Chemical stability within the components to maintain ion transport.

5) Stabilized structure after being treated to high temperature.

Oxide type garnet $Li_7La_3Zr_2O_{12}$ (LLZO) has these above listed properties up to some extent. It has a high electrochemical potential window, comparable ionic conductivity, and stability against lithium metal anode [25]. Because of these benefits, LLZO can be treated as a solid electrolyte for upcoming ASSB's. Initially, garnet structure LLZO has a tetragonal phase ($I4_1/acd$) at room temperature. After heat treatment, this phase gets modified into the cubic phase ($Ia\bar{3}d$) which has a higher ionic conductivity compared to tetragonal phase. But, the main challenge is stabilizing this cubic phase at room temperature, which only occurs at higher temperatures. Such high temperature procedure leads to loss of Li which increases interfacial resistance and minimizes the ionic conductivity. This temperature range is also not desirable for the assembly of ASSB's. To overcome this issue, many ways have been developed, such as the addition of additives to minimize the sintering temperature and the substitution of supervalent cations. Among this, adding cations in the main lattice of LLZO brings a significant change. This substitution can occur at any of the Li, La, and Zr sites. With substitution at Li lattice, the content of Li reduces and increases the Li vacancies, which not only helps to stabilize the cubic phase but also increases the ionic conductivity by creating Li-ion conduction pathways [17]. Substitution of supervalent cations also acts as a sintering aid, lowering the essential temperature range for cubic phase formation. This phenomenon was first noticed when the accidental contamination of Al happened from the alumina crucible to the LLZO. This stabilizes the cubic phase, which is then intentionally substituted in LLZO to study its effect on ionic conductivity. And it was found that the formed structure was denser than the earlier one, with stabilized cubic phase and increased conductivity [26].

Thus after this discovery, to get stabilized cubic phase with increased ionic conductivity, the substitution of many supervalent cations like $Al^{3+}$, $Ga^{3+}$, $Be^{3+}$, $B^{3+}$, $Fe^{3+}$, $Zn^{3+}$, $Ge^{4+}$ at the Li site has been studied. On the other hand, substitution at La and Zr sites has also been examined with various cations like $Ba^{2+}$, $Nb^{5+,}$ and $Ta^{5+,}$ respectively [27–29]. Along with single-site substitution, many investigations also proved the enhancement in conductivity by simultaneous substitution method where $Rb^{3+}$ and $Ca^{2+}$ were substituted at the La site, respectively and $Ta^{5+}$ substituted at the Zr site [30,31]. Substitution at the Li site with $Ga^{3+}$ and $Al^{3+}$ also significantly affects structural and electrical properties [32]. The co-doping of $Ga^{3+}$ and $Y^{3+}$ at Li and La site has also proved the stability of this electrolyte [33]. This co-doping method generates the synergistic effect on Li-ion conduction in garnet LLZO. In between all this study, $Ta^{5+}$ doped LLZO indeed shows the desirable results. The previously investigated reports regarding the methods and their corresponding ionic conductivity for $Ta^{5+}$ doped LLZO have been mentioned in Table no 1. It indicates that $Ta^{5+}$ substituted LLZO shows excellent results due to its ability to stabilize the cubic phase and its relative stability to Li. It does not obstruct the conduction path of Li ions. Thus, it is interesting to study more about the co-doping effect of $Ta^{5+}$ with some other supervalent cations, which helps in increasing the ionic conductivity of garnet LLZO. Higher room temperature conductivity by co-doping at the Li and Zr sites have been reported earlier[28].

Thus, in this present work, the effect of co-doping of $Ge^{4+}$ at the Li site and $Ta^{5+}$ at the Zr site in LLZO has been studied. Ge4+ has been selected here because it can form cubic phase at lower temperatures. Thus, a much lower sintering temperature has been utilized for the synthesis due to the co-doping effect. The concentration of $Ge^{4+}$ has been kept fixed to optimize the Li content. For higher Li ion conductivity, it is generally believed that the minimum Li content should be in the range of 6.4-6.6 atoms per formula unit [28]. With the fixed specific amount of $Ge^{4+}$, the concentration of $Ta^{5+}$ has been varied between 0.15 – 0.45. The reported optimized comparisons with maximum ionic conductivity are given in table 1.

## 2. Experimental

### *2.1. Sample Preparation*

The $Li_{6.8-y}Ge_{0.05}La_3Zr_{2-y}Ta_yO_{12}$ (y = 0, 0.15, 0.25, 0.35, 0.45) series has been prepared by solid-state reaction method using precursors: $Li_2CO_3$ (Merck, >99.9%), $La_2O_3$, $ZrO_2$, $GeO_2$ and $Ta_2O_5$ (Sigma Aldrich, >99.99%). All the stoichiometrically weighted chemicals were hand mixed in an agate mortar. To overcome the loss of Li during sintering, 10% excess $Li_2CO_3$ was mixed. The powder was then kept in a muffle furnace for calcination at $900^0$ C for 8 hr using an alumina crucible. The calcined powder, after cooling to room temperature was crushed into fine powder. The pellets are made using diaset of 1.5 mm thickness and 10 mm diameter with uniaxial pressure of 4 tons per $cm^2$ using hydraulic press. The formed pellets then covered with mother powder and kept in the furnace for sintering at $1050^0$ C for 7.30 h, covered with an alumina lid in an ambient atmosphere. All the prepared ceramic samples are abbreviated as 0 Ta, 0.15 Ta, 0.25 Ta, 0.35 Ta, and 0.45 Ta for respective values of y ranging from 0-0.45, respectively.

### *2.2. Sample Characterizations*

The formed pellets after sintering at $1050^0$ C were crushed into fine powder for phase identification, and the fragments were then investigated using X-ray diffraction (RIGAKU diffractometer) with Cu-kα radiation having a wavelength of 1.54 $A^0$ as a radiation source. The data were collected at a scan speed of $2^0$/min and with a step size of 0.02 degrees in the $10^0$–$80^0$ range. Using the K-15 Classic (K-Roy) instrument and toluene as an immersion medium, the densities were caluculated by using the Archimedes' principle. The microstructural investigation and compositional study were carried out using scanning electron microscopy (JSM-7600 F/JEOL) with an enegy dispersive spectroscope attached to it. A novocontrol impedance analyzer was used in the frequency range from 20 Hz–20 KHz at various temperatures from room temperature to $150^0$ C to obtain the ionic conductivity and activation energy. Both the faces of sintered pellet has been coated with silver paste to maintanin the ohmic contact with the silver electrodes. These electrodes serve as blocking electrodes for both the AC and DC conductivity. KEITHLEY 6512 programmed electrometer was used to calculate ionic transport number. The constant voltage of 1 V was applied to the silver electrodes, and the current was measured in equal time intervals.

## 3. Results and Discussion

### *3.1. X-ray Diffraction Study*

The series $Li_{6.8-y}Ge_{0.05}La_3Zr_{2-y}Ta_yO_{12}$, synthesized using a solid-state reaction method with y = 0-0.45, shows the dominance of a single cubic phase (space group – Ia-3d). Fig.(1-a) shows the X-ray diffraction patterns of all the samples with respective (h k l) planes. All the peaks associated with the high conducting cubic phase have been matched with the cubic garnet structure $Li_5La_3Nb_2O_{12}$ having JCPDS file no. 45.0109 (represented as a black vertical line in the graph). The graph of sample without Ta, shows the splitting of peaks (In fig. 1-b), whereas all the other ceramics show neither splitting nor external impurity peak. This means Ta helps the structure to form and stabilize the cubic phase at room temperature. This result is in good accordance with the results reported by J.L. Allen et al.[26]. From the graph, it is clear that, with the increase in Ta concentration, the intensity and sharpness have been increased, resulting in high crystallinity. Cao Yang et al. [25] suggested that for the formation of a cubic phase with y = 0.25 Ta, the $1150^0$ C – $1225^0$ C temperature range is suitable with a sintering time of almost 36h. Also, even after the addition of LIOH with the sintering temperature of $1100^0$ C for 16 h, the pyrochlore phase $La_2Zr_2O_7$ has developed [38]. But in this study, the cubic phase has been obtained with the sintering temperature of $1050^0$ C and time of 7.30 h. This can be attributed to the insertion of Ta and the substitution of Ge in LLZO, which acts as a sintering aid in ceramics [28]. The successful insertion of Ta into LLZO can be seen in Fig. (1-b). The shifting of peaks towards higher angles indicates a decrease in lattice parameters. And this can be due to the smaller ionic radius of $Ta^{5+}$ (0.69 $A^0$) than $Zr^{4+}$ (0.78 $A^0$). The average crystallite size for 0.25 Ta was calculated using the Debye-Scherrer formula, and it is found to be 31.70 nm.

### *3.2. Density Measurement*

The densitities of all the ceramics have been calculated based on physical parameters using the Archimedes' principle. Toluene was used as an immersion medium. The values of density of all samples with their respective relative densities are mentioned in table 2. From the table, it is clear that the density increases with the increase in Ta concentration. In some previous studies, where the samples were prepared by the solid-state reaction method and sol gel method, the highest density was achieved by adding the external additives with the hot pressing and high temperature sintering method for several hours [25,39,40]. However, in this study, the highest density occurs for 0.25 Ta, which is 4.77g.cm$^{-3}$. The relative density is calculated using the therotical density value obtained from reitveild refinement. This density value is greater than the earlier reported density value for same Ta content [25]. This can be due to the co-doping of $Ge^{4+}$ and $Ta^{5+}$ in the lattice of garnet LLZO [41]. This might help in the formation of compact ceramic electrolyte as it helps in grain growth, which eventually combines the neighboring grains [28]. This result can be seen from the SEM micrograph of 0.25 Ta ceramic in Fig.2(a).

### *3.3. Morphological Study*

Fig. 2(a) represents the SEM micrograph of 0.25 Ta ceramic. The grain boundary resistance may arise due to the high porosity in solid ceramic electrolytes. This eventually leads to reduced Li-ion conductivity. From the figure, it can be clearly seen that all the neighboring grains are very well connected, forming a dense structure. This result is in aggrement with the maximum density value of 0.25 Ta ceramic which is given in table 2. Previously reported results for 0.25 Ta shows countable pores, which might be due to the volatilization of $Li^{+}$ due to high sintering temperature[25]. Thus the result obtained in current study can be attributed to the co-doing of $Ge^{4+}$ with $Ta^{5+}$ in LLZO [41]. In this study, there is an advantage of large grain size with negligible voids which provides the conduction pathways for $Li^{+}$ ions, giving the maximum conductivity. The histogram in fig. 2 (b) represents the average particle size calculated using Image-J software. The average particle size is 8.98 µm. This obtained result of grain size also supports the fact that it is not necessary to get a lower density value for a higher grain size [34], which can be seen from the density values mentioned in table 2. Thus, 0.25 Ta ceramic has the advantage of maximum density with a large grain size.

Fig. 3 shows the elemental mapping of the constituent elements of 0.25 Ta sample. From the figure, it can be observed that, all the elements viz. La, Zr, Ge, and Ta are present within the sample and are evenly spread all over the ceramic. This distribution suggest that, the Ta and Ge are incorporated within the LLZO garnet. Here, Li cannot be detected due to its low energy. The presence of Ta surrounding the grain boundary is also confirmed by the XRD result in which the peaks are shifted towards the higher value of theta.

### *3.4. AC conductivity Studies*

#### *3.4.1.Impedance plots*

The complex impedance plots at room temperature for the series $Li_{6.8-y}Ge_{0.05}La_3Zr_{2-y}Ta_yO_{12}$ with y = 0 – 0.45 have been shown in fig 4(a). The impedance can be measured by the intercept made by the semicircle on the real z-axis. The semicircle in the higher frequency region and the tail in the lower frequency region suggest that the conduction is primarily ionic [42]. However, the single semicircle in higher frequency region is the combination of semicircles due to grain and grain boundary, and the appearance of the tail is due to the ion-blocking nature of the Ag electrodes [43]. The ionic conductivity has been calculated using the formula $\sigma_{total} = t/RA$ ,where, $\sigma_{total}$ is the ionic conductivity, $t$ is the thickness of the sample , $R$ is the resistance offered by the sample, and $A$ is the area of the electrode. From fig.4 (a), it can be observed that the sample without Ta (0 Ta) possesses the highest resistance. But, with the increase in the Ta content, the semicircles shifted towards the lower resistance value. The minimum value of resistance is offered by 0.25 Ta, which has the maximum ionic conductivity of 6.61 x $10^{-4}$ S/cm at $25^0$C among all the samples. Fig. 4 (b) shows the fitted spectra for 0.25 Ta, where $R_1$ and $R_2$ are the grain and grain boundary resistance. The obtained result for 0.25 Ta is supported by the previously mentioned highest density value and dense microstructure. The co-doping of $Ge^{4+}$ with $Ta^{5+}$ helps in reducing the resistance by increasing Li vacancies, promoting Li ion migration pathways, and helping to optimize the total Li content in the sample [28]. But beyond 0.25 Ta, further increase in Ta content results in decrease in ionic conductivity. This may be due to the excess of Ta, which reduces the required Li content and affects its conducting ability [31]. Hence, from the obtained result, it can be concluded that, for $Li_{6.8-y}Ge_{0.05}La_3Zr_{2-y}Ta_yO_{12}$, y = 0.25 is the optimum Ta content. This result is also well supported by earlier reporting by Yang et al.[25]. To check the stability of ionic conductivity offered by the 0.25 Ta ceramic, it was further exposed to air for 5 months. The ionic conductivity after exposure was calculated from the graph shown in fig.4 (c). And Fig.4 (d) shows the fitted spectra of the air-exposed 0.25 Ta ceramic. From the fig.4 (c), it can be seen that, as compared to the original 0.25 Ta sample, the semicircle for the 0.25 Ta sample after exposed to air, slightly shifted towards the higher resistance region, giving the total conductivity value of 5.95 x $10^{-4}$ S/cm. This conductivity value is slightly less than the original conductivity value 6.61 x $10^{-4}$ S/cm. This result confirms the stability of the 0.25 Ta ceramic owing to co-doing of Ge and Ta in LLZO.

### *3.4.2. Arrhenius plots*

Arrhenius plots for the series $Li_{6.8-y}Ge_{0.05}La_3Zr_{2-y}Ta_yO_{12}$ (y=0-0.45) are shown in fig.5 (a). The measurements have been taken in the $25^0$ C – $150^0$ C temperature range. The activation energy is calculated using the Arrhenius equation , $\sigma(T) = \sigma_0 exp(-E_a/K_B T)$. Here, $\sigma$ is the conductivity, $\sigma_0$ is the pre-exponential factor, $E_a$ is the activation energy, $K_B$ is the Boltzmann constant, and $T$ is the temperature in Kelvin. The maximum conductivity of 6.61 x $10^{-4}$ S/cm is obtained for 0.25 Ta ceramic having minimum activation energy of 0.25 eV. The conductivity and activation energy values have been listed in table 3. Also, the variation of ionic conductivity and activation energy as a function of varying content of Ta is shown in fig.5 (b). The obtained value of activation energy for 0.25 Ta slightly differs from the result reported by Xingxing Zhang et al. [38] for the same Ta content usingLi rich atmosphere. But here, the addition of $Ta^{5+}$ with $Ge^{4+}$ , without excessive Li content, gives a similar result with lower activation energy than other reported studies [38]. The increased conductivity value with minimum activation enrgy can also attributed to $Ta^{5+}$ doping which helps in stabilizing cubic phase with increase in Li vacancy and distortion with minimum content (<0.6 pfu) [44].

### *3.5. DC conductivity study*

Fig. 6 shows the DC conductivity graph for the 0.25 Ta sample. The ionic transport number measurement has been done to confirm the predominance of ionic conductivity over electronic conductivity. Here, silver electrodes were used where a 0.25 Ta ceramic was placed. The constant voltage of 1 V was applied to the ceramic, and the current through the ceramic was measured for equal intervals of time. After some time, the conductivity remains unchanged, and it is believed to be due to electrons [45]. The ionic transport number is calculated using the formula as $t_i = (\sigma_{total} - \sigma_e)/\, \sigma_{0total}$. For 0.25 Ta ceramic, the ionic transport number was found to be greater than 0.999. This confirms the dominant ionic conduction in the prepared sample[44].

## 4. Conclusions

The series $Li_{6.8-y}Ge_{0.05}La_3Zr_{2-y}Ta_yO_{12}$ (y=0-0.45) has been prepared by conventional solid-state reaction method. All the ceramic samples show a highly conductive cubic phase. The 0.25 Ta ceramic has a maximum density with dense microstructure.The elemental mapping shows the uniform distribution of elements in 0.25 Ta ceramic. There is an increase in the ionic conductivity by two orders of magnitude with 0.25 Ta content. The maximum conductivity value was 6.61 x $10^{-4}$ S/cm with a minimum activation energy of 0.25 eV for 0.25 Ta ceramic. This has been attributed to the co-doping effect of Ge and Ta in LLZO as Ge acts as a sintering aid and Ta helps in the stabilization of cubic phase. The 0.25 Ta ceramic is stable after air exposure for 5 months. The highest conducting 0.25 Ta ceramic is stable in the ambient atmosphere. The DC conductivity measurements confirmed the predominant nature of ionic conductivity. Thus 0.25 Ta ceramic sample is a promising candidate as solid electrolyte for all-solid-state lithium-ion batteries.

**Acknowledgment**

One of the authors wishes to acknowledge VNIT, Nagpur, for offering a Ph.D. fellowship. The Department of Physics at VNIT, Nagpur, provided the XRD facility, and the author is thankful for the support of DST FIST project number SR/FST/PSI/2017/5(C).

**Funding**

This research did not receive any specific grant from funding agencies in the public, commercial, or not-for-profit sector.

**Figures and caption:**

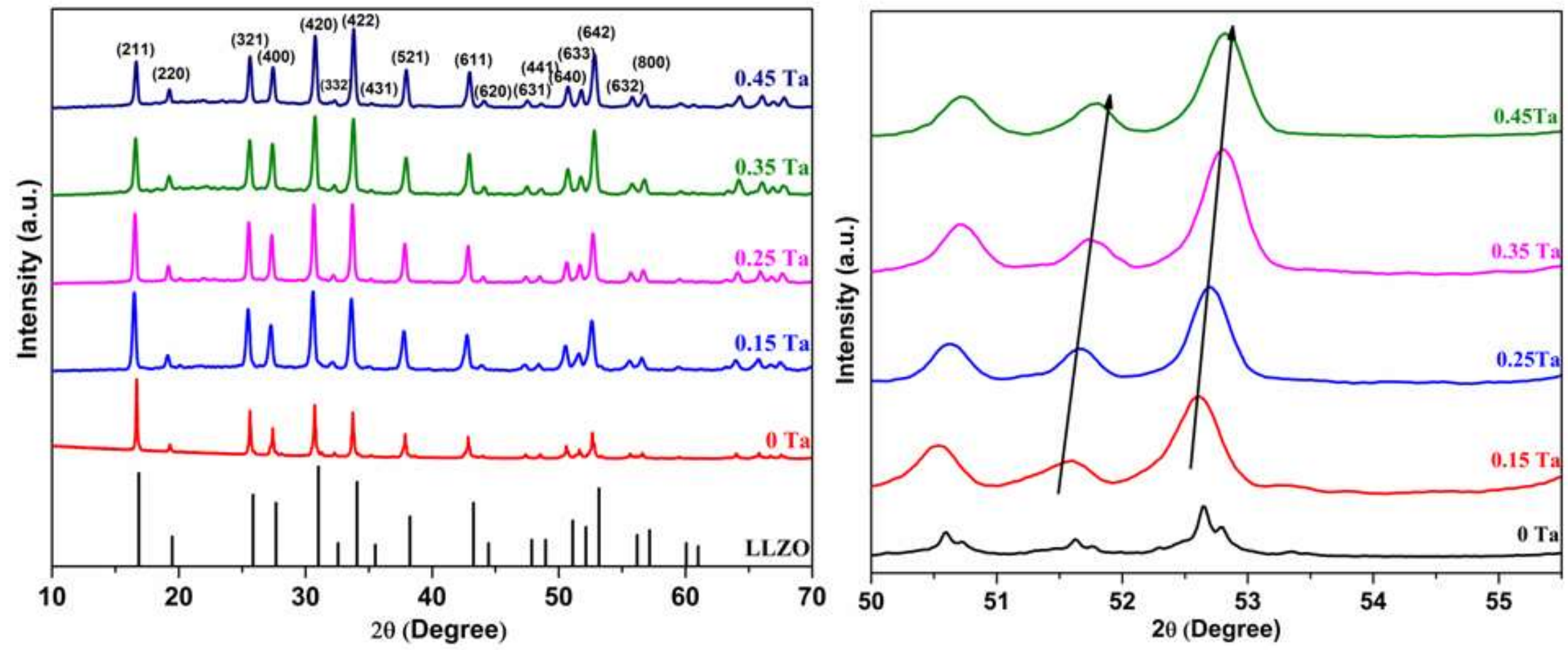

**(a)** **(b)**

**Fig.1.** **(a)** XRD patterns and **(b)** Shifting of peaks of $Li_{6.8-y}Ge_{0.05}La_3Zr_{2-y}Ta_yO_{12}$ with y ranging from 0 – 0.45.

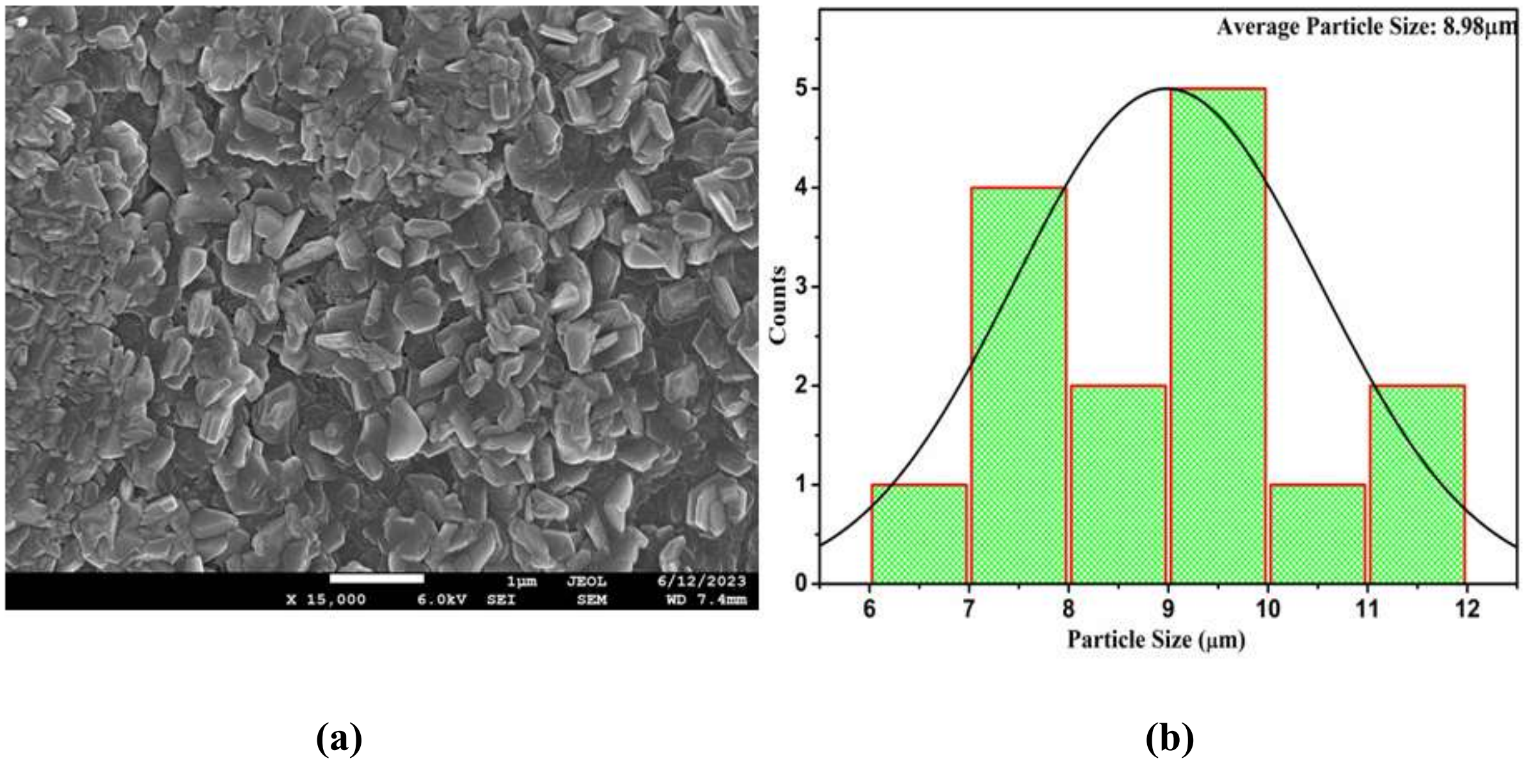

**(a)** **(b)**

**Fig.2.** **(a)** SEM image of 0.25 Ta and **(b)** Average particle size distribution of 0.25 Ta.

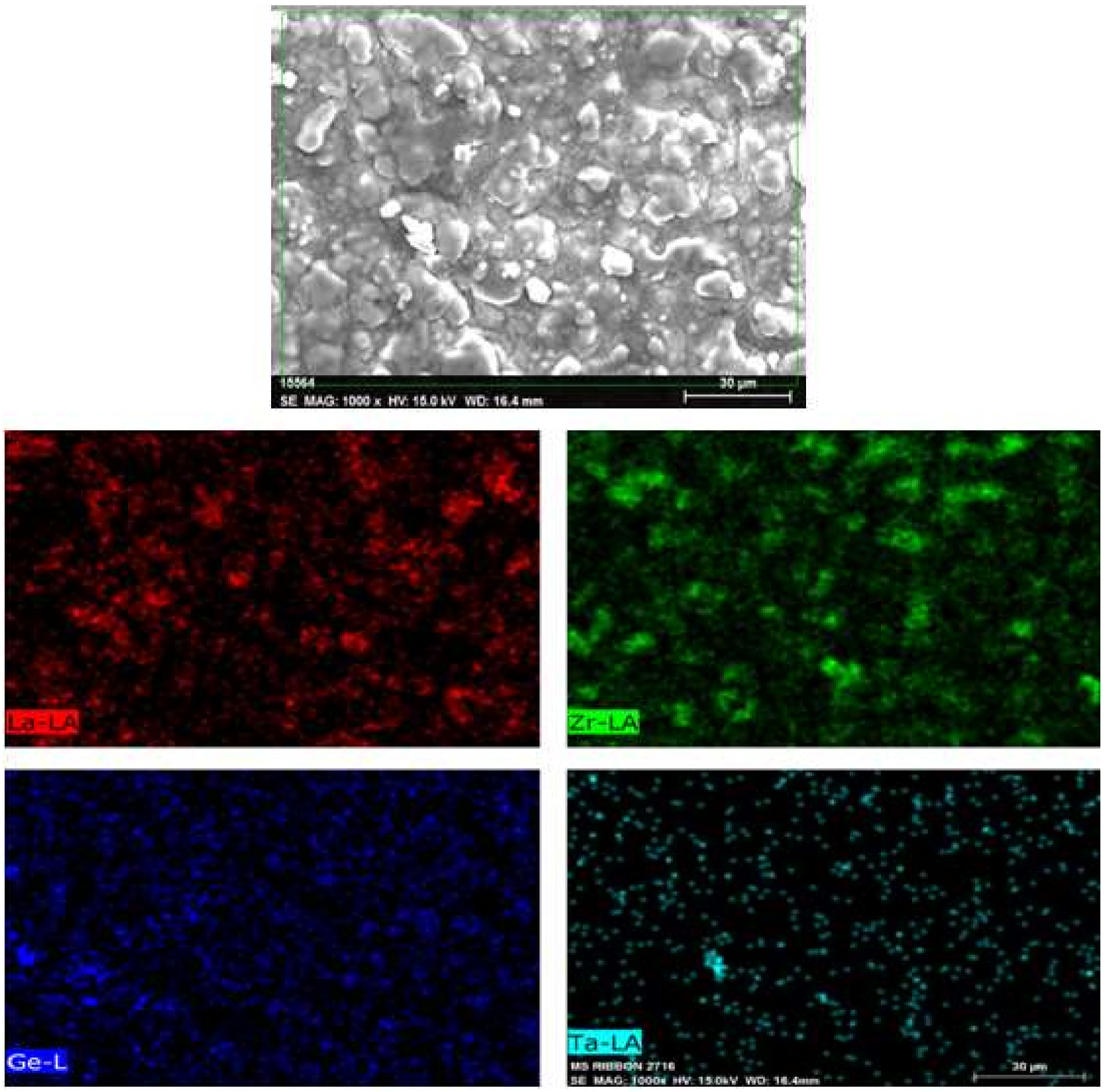

**Fig.3.**Elemental mapping of 0.25 Ta ceramic.

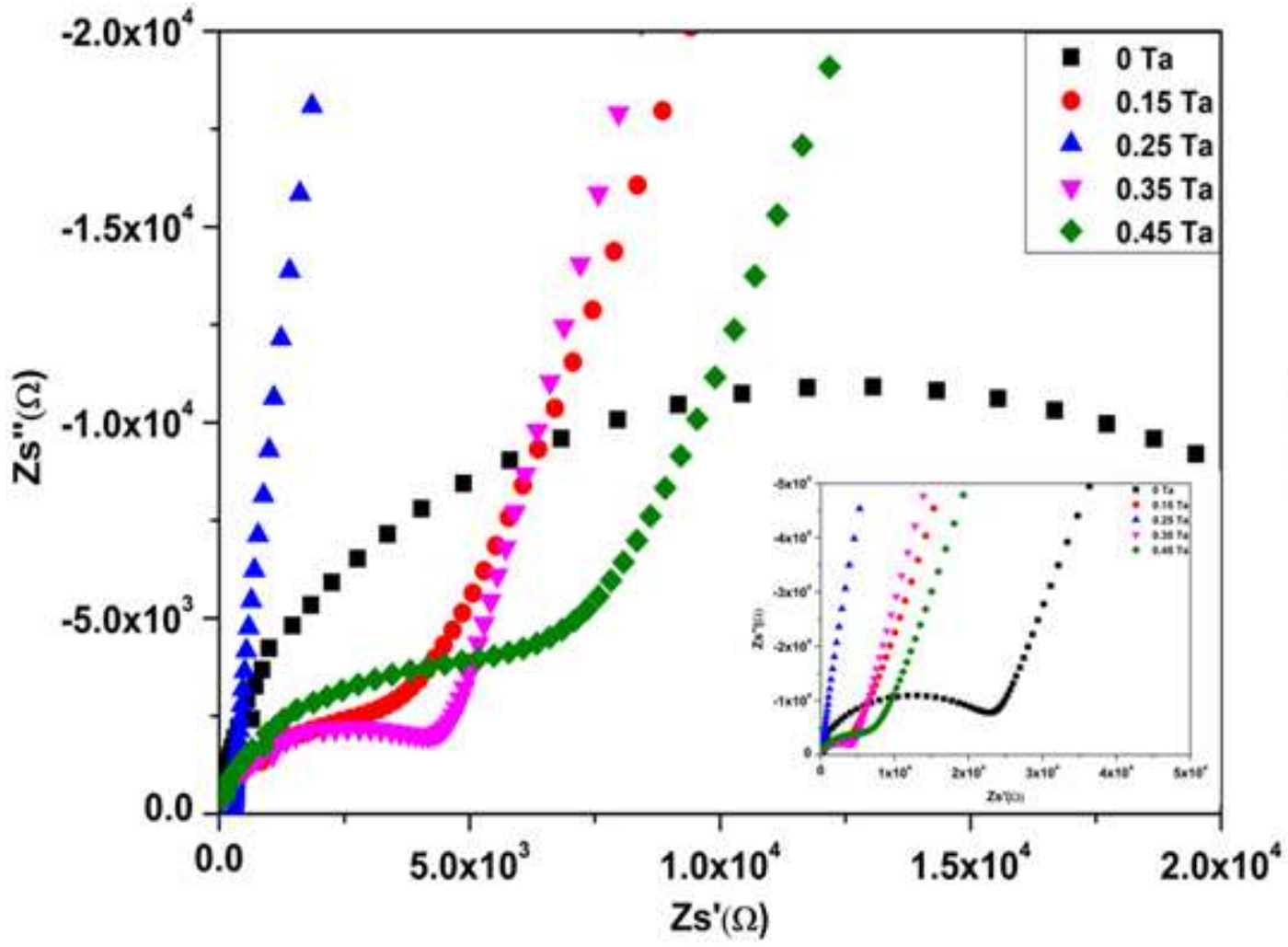

**(a)**

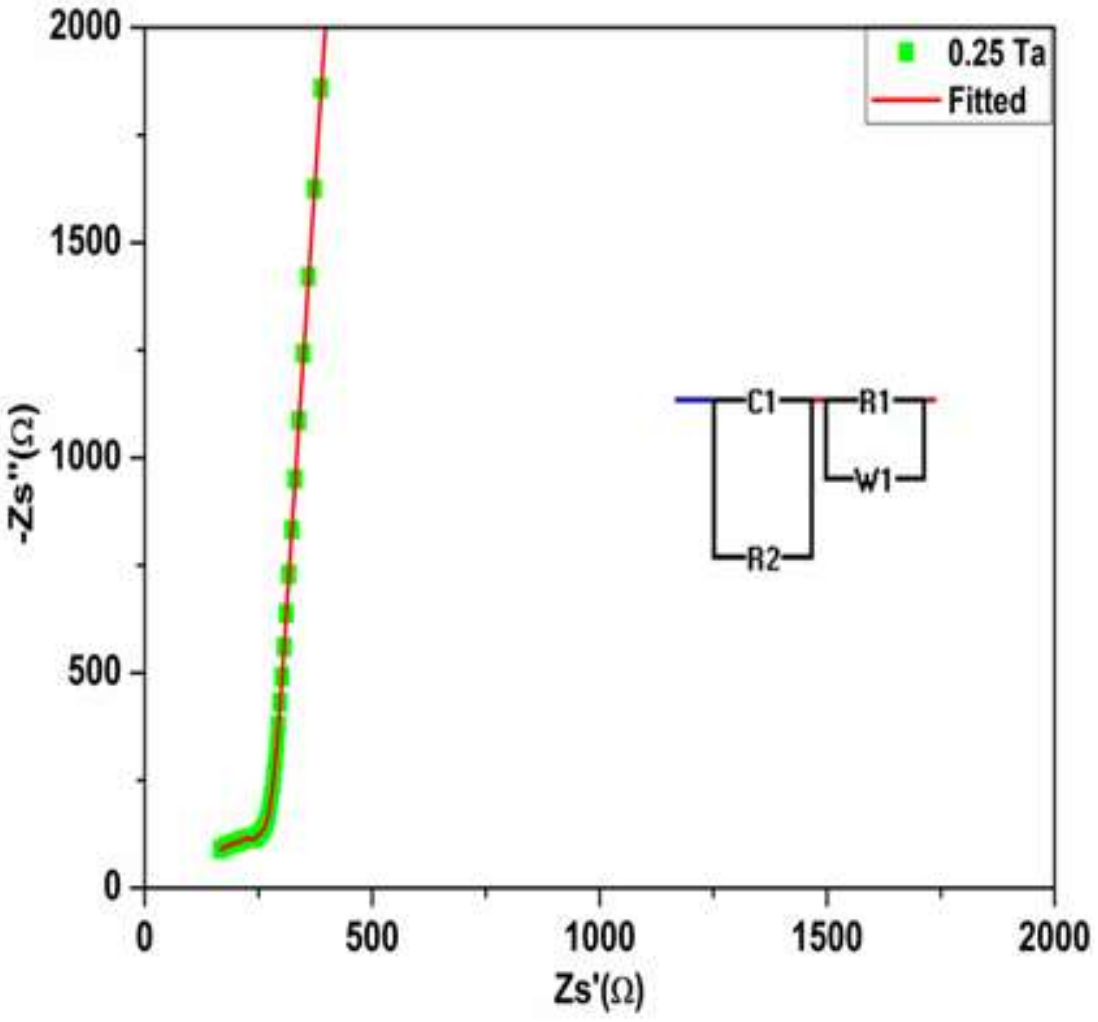

**(b)**

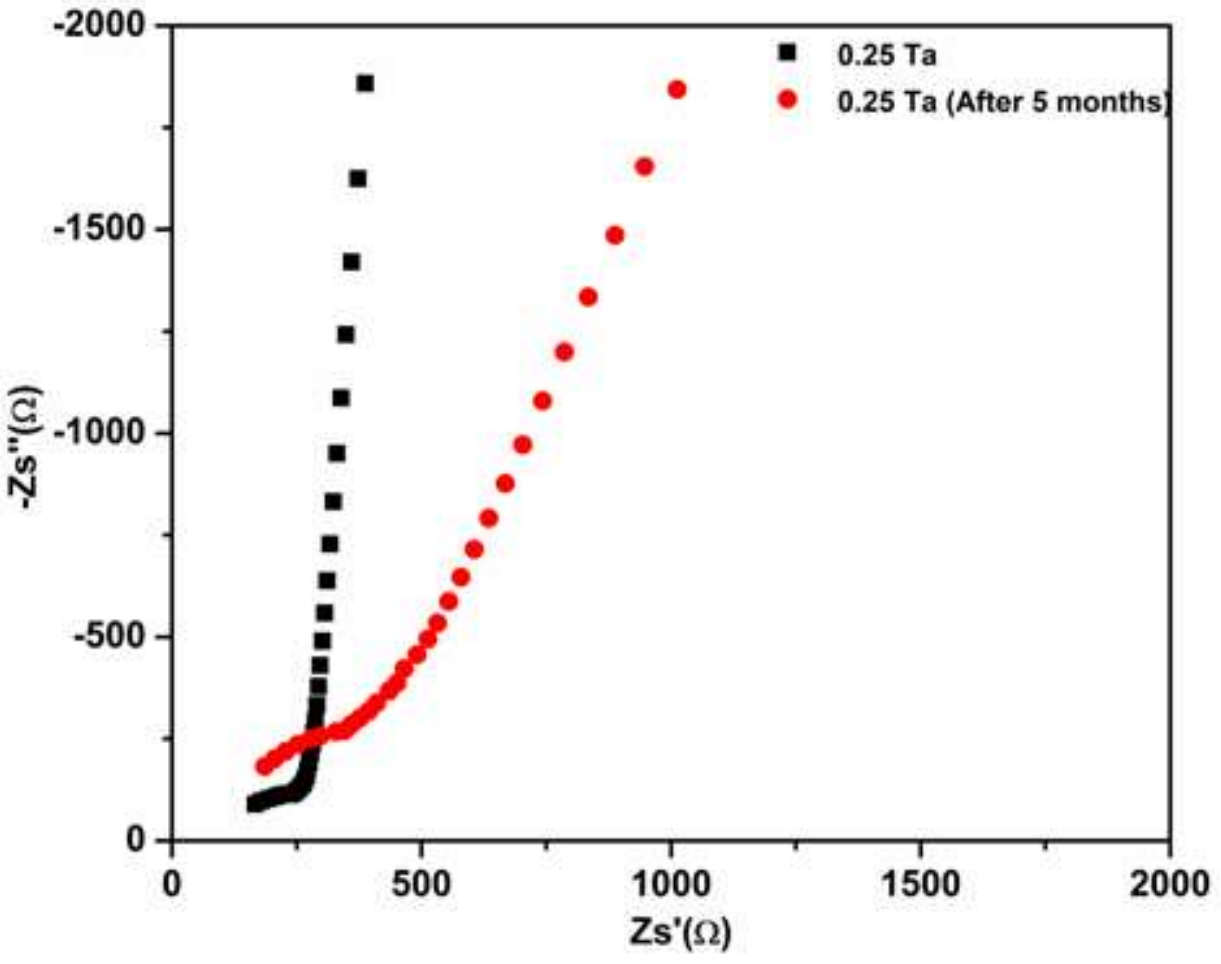

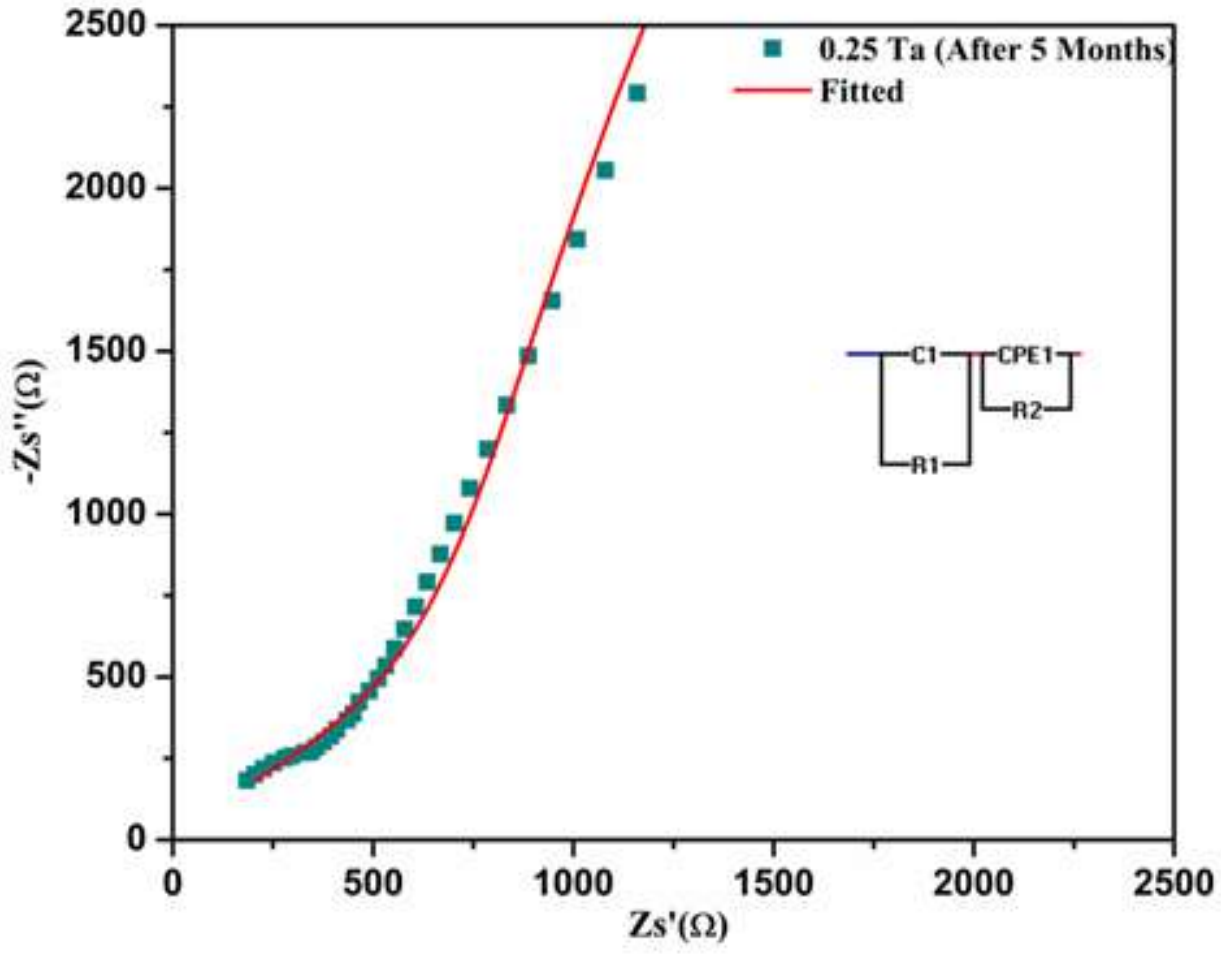

**(c)**

**(d)**

**Fig.4. (a)** Nyquist plots for $Li_{6.8-y}Ge_{0.05}La_3Zr_{2-y}Ta_yO_{12}$ with y ranging from 0-0.45. **(b)** Fitted plot for $Li_{6.8-y}Ge_{0.05}La_3Zr_{2-y}Ta_yO_{12}$ with y = 0.25 (0.25 Ta) at $25^0$C. **(c)** Nyquist plot of 0.25Ta ceramic after air exposure for 5 months. **(d)** Fitted graph of 0.25 Ta ceramic after air exposure..

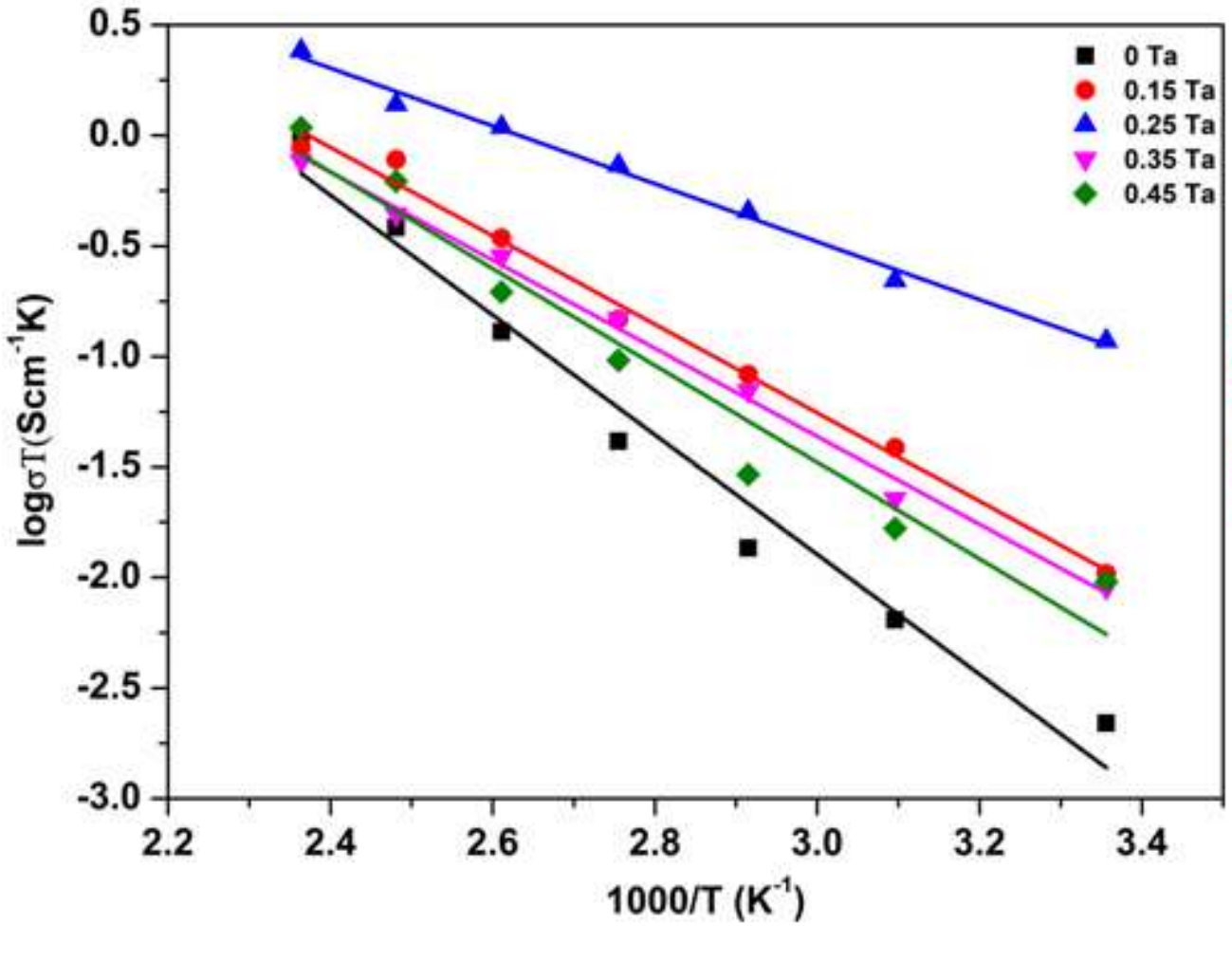

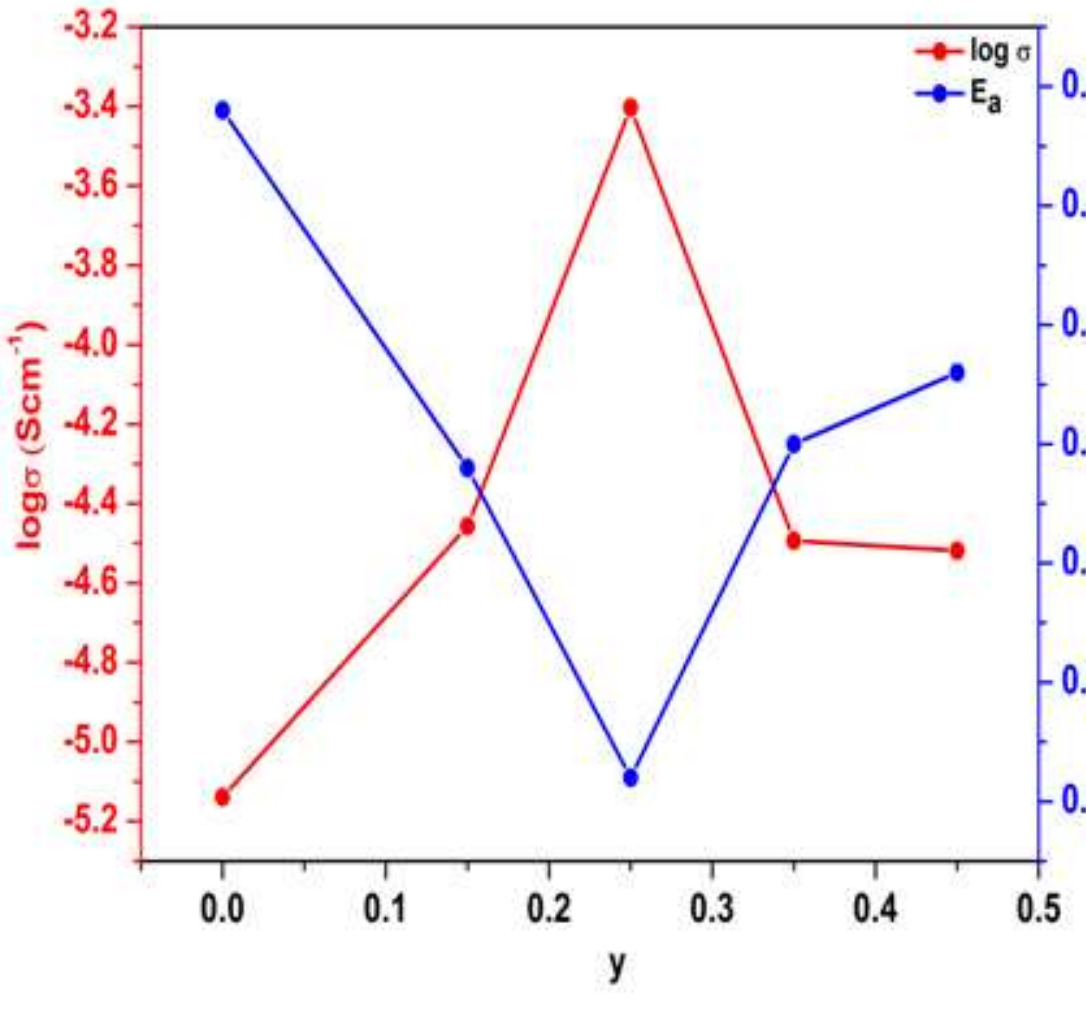

**(a)**

**(b)**

**Fig.5. (a)** Arrhenius plots for $Li_{6.8-y}Ge_{0.05}La_3Zr_{2-y}Ta_yO_{12}$ with y = 0 -0.4

**(b)** Variation in conductivity and activation energy with content of Ta (y).

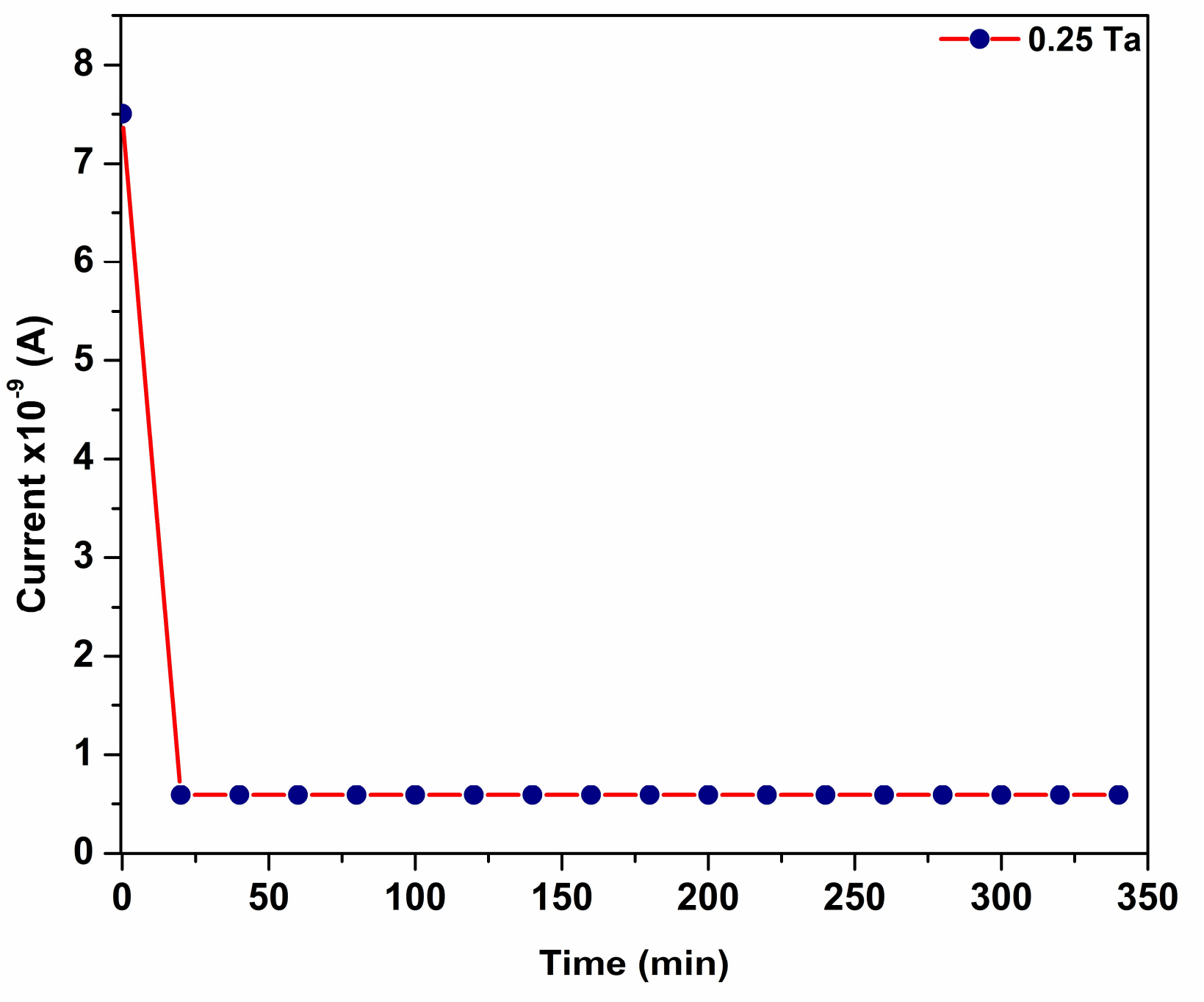

**Fig.6.** DC conductivity measurement with respect to time and current.

**Tables and caption:**

**Table 1:** Ionic conductivity values at room temperature ($25^0$C) of $Ta^{5+}$ doped LLZO with various additives at different sintering temperature and time.

| Composition | Additives | Synthesis Process | Sintering Temperature & Time | Ionic Conductivity (S/cm) | Reference |
|---|---|---|---|---|---|
| **Ta-LLZO** | - | Two-step synthesis | $1175^0$C/12h | $9.58 \times 10^{-5}$ | [34] |
| **LLZTO** | LCBO | Solid state reaction | $1000^0$C/15h | $1 \times 10^{-4}$ | [35] |
| **LLZTO** | CuO | Solid-state reactive sintering | $1100^0$C/10-30 h | $2.99 \times 10^{-4}$ | [36] |
| **$Li_{6.5}La_3Zr_{1.5}Ta_{0.5}O_{12}$** | $Li_2CO_3$ | SolidState Sintering | $1140^0$C/16h | $3.16 \times 10^{-4}$ | [36] |
| **$Li_{6.4}La_3Zr_{1.4}Ta_{0.6}O_{12}$** | $Li_4SiO_4$ | Solid state reaction | $1200^0$C/24h | $3.21 \times 10^{-4}$ | [36] |
| **$Li_{6.4}La_3Zr_{1.4}Ta_{0.6}O_{12}$** | MgO | Solid state reaction | $1180^0$C/5h | $3.35 \times 10^{-4}$ | [36] |
| **$Li_{6.4}La_3Zr_{1.4}Ta_{0.6}O_{12}$** | - | Sol-gel method | $1150^0$C/1h | $1.4 \times 10^{-4}$ | [37] |
| $Li_{6.55}Ge_{0.05}La_3Zr_{1.75}Ta_{0.25}O_{12}$ | - | **Solid Sate Reaction** | **$1050^0$C/7.30h** | **$6.61 \times 10^{-4}$** | **Present work** |

**Table 2:** Density and relative density values of $Li_{6.8-y}Ge_{0.05}La_3Zr_{2-y}Ta_yO_{12}$ (y = 0-0.45).

| Ta content | Density ($g.cm^{-3}$) | Relative Density (%) |
|---|---|---|
| 0 | 4.34 | 84.96 |
| 0.15 | 4.60 | 90.05 |
| **0.25** | **4.77** | **93.38** |
| 0.35 | 4.65 | 91.03 |
| 0.45 | 4.62 | 90.44 |

**Table 3:** Ionic conductivity and activation energy values of $Li_{6.8-y}Ge_{0.05}La_3Zr_{2-y}Ta_yO_{12}$ (y = 0-0.45).

| Content of Ta (y) | Ionic Conductivity (S/cm) | Activation Energy (eV) |
|---|---|---|
| 0 | $7.64 \times 10^{-6}$ | 0.56 |
| 0.15 | $3.50 \times 10^{-5}$ | 0.39 |
| **0.25** | **$6.61 \times 10^{-4}$** | **0.25** |
| 0.35 | $3.20 \times 10^{-5}$ | 0.40 |
| 0.45 | $3.02 \times 10^{-5}$ | 0.43 |